# SYSTEM LEVEL NUMERICAL ANALYSIS OF A MONTE CARLO SIMULATION OF THE E. COLI CHEMOTAXIS

C. I. Siettos

School of Applied Mathematics and Physical Sciences, National Technical University of Athens, GR 157 80, Greece

*ABSTRACT*

Over the past few years it has been demonstrated that "coarse timesteppers" establish a link between traditional numerical analysis and microscopic/ stochastic simulation. The underlying assumption of the associated "lift-run-restrict-estimate" procedure is that macroscopic models exist and close in terms of a few governing moments of microscopically evolving distributions, but they are unavailable in closed form. This leads to a system identification based computational approach that sidesteps the necessity of deriving *explicit* closures. Two-level codes are constructed; the "outer" code performs macroscopic, continuum level numerical tasks, while the "inner" code estimates –through appropriately initialized "bursts" of microscopic simulation- the quantities required for continuum numerics. Such quantities include residuals, time derivatives, and the action of coarse slow Jacobians. We demonstrate how these "coarse timesteppers" can be applied to perform equation-free computations of a kinetic Monte Carlo simulation of *E. coli* chemotaxis. Coarse-grained contraction mappings, system level stability analysis as well as acceleration of the direct simulation, are enabled through this computational multiscale "enabling technology".

*Keywords: Bio-mechanics, Chemotaxis, Microscopic Models, Monte-Carlo simulations, Multi-scale Computations, Equation-Free approach*

## 1. Introduction

Chemotaxis, the process by which cells change their speed and/or orientate in space responding to environmental chemical gradients, has been thoroughly studied as it constitutes one of the basic survival and growth mechanisms of many micro-organisms. Revealing and understanding these mechanisms that pertain the function of this biological directional "kinesis" is of outmost importance in several fields of biology, biomechanics and medicine including homeostasis [Yi et al., 2000; Ito et al., 2004], ocean ecology[Stocker et al., 2008], biologically-inspired robotics [Long et al., 2004] and orthopedics [Lend et al., 1995; Poitout, 2004]. The chemotaxis pathway and biomechanics of E.coli have been well investigated through experiments [Macnab and Koshland, 1972, Brown and Berg,1974; Berg, 1975; Parkinson, 1993; Spiro et al., 2997; Waldo et al., 1999; Cluzel et al., 2000] providing useful insights also for other chemotactic species. A recent study provided that more than half of the bacteria have chemotactic ability [Wuichet and Zhulin, 2010].

The extensive study of E-coli has allowed the development of a many dynamical models ranging from the macro [Patlak, 1953; Keller and Segel, 1971a,b; Lapidus and Schiller, 1974; Biler, 1998; Dolak and Schmeiser, 2005 ] to the mechanistic-based micro-scale incorporating biological information even at the molecular inner-cell level [Alt, 1980; Rivero et al., 1989; Othmer and Stevens, 1997; Stevens, 2000; Hillen and Othmer, 2000; Painter et al., 2000; ]. For a detailed review of these models please refer to [Horstmann, 2003; Tindall et al., 2008; Hiller and Paiter, 2009]. No doupt, these mathematical models are invaluable tools that can help shedding light on still unknown issues regarding not only chemotaxis but also many other "adaptive" biochemical and genetic networks as well as human health [Wu, 2005; Wuichet and Zhulin, 2010] and life-cycle [Friedrich and Ju licher, 2007].

However, current modeling practice in the field is often characterized by the lack of accurate macroscopic deterministic models that quantitatively reflect the dynamic behavior of the systems modeled. Due to effectively stochastic nature and nonlinear complexity of the biological dynamics, state equations resulting at the macroscopic level from the appropriate balances are simply not available, or overwhelming difficult to derive.

For example, under certain assumptions the dynamics at the macroscopic level can be represented by a Partial-Differential Equation (a Fokker-Planck) describing the evolution of the probability density, say, $\rho(x)$ of the bacteria in space [Snitzer, 1993] :

$$\frac{\partial \rho(x)}{\partial t} = \wp(x, \rho, p) \tag{1}$$

where $\wp$ is the partial differential operator and $p$ denotes the vector containing other "internal" variables. If cell devision or death is ruled out, the above equation reads [Othmer and Schaap, 1998]:

$$\frac{\partial \rho(x)}{\partial t} = \nabla \cdot (D\nabla\rho - \rho u) \tag{2}$$

where $D$ is the diffusion coefficient and $u$ is the chemotactic velocity.

The above equation is not in a closed form as $u$ is not explicitly related to $\rho$. Over the years, various approximations have been proposed that express $u$ in terms of other internal variables such as the gradient of the concentration of the attractant (or repellant). For example, a most common approximation, as extracted from kinetic theory is given by the Patlak-Keller/Segel relation reading [Patlak, 1953; Keller and Segel, 1971a,b]:

$$u = \chi \nabla c \tag{3}$$

where $\chi$ denotes the chemotactic sensitivity, $c$ is the concentration of the chemo-attractant (or repellant). The above closure is derived assuming that the cells emit immediately the chemical chemoattractants which are then diffused.

Many other approaches for the calculation of the chemotactic sensitivity or the chemotactic velocity expressed in terms of other internal variables have been proposed (see the discussion in [Othmer and Schaap, 1998]. Yet, such closure relations are based on approximations and assumptions that introduce certain biases in the analysis. This prevents the performance of important computational tasks such as stability and bifurcation analysis, control and optimization, which rely on the availability of models, lumped or distributed, in terms of macroscopic (coarse) variables.

On the other hand, good models of the underlying physics can be written at the microscopic/cellular level, and the modeling is performed via stochastic simulators (e.g. Monte Carlo, Brownian dynamics, Lattice-Boltzmann, Markov chains). However, at this fine biological level of representation, standard system analysis and design algorithms cannot be used efficient to handle the given detailed information.

Towards this aim, it has been shown that "coarse timesteppers" establish a link between traditional numerical analysis and microscopic simulation [Gear et al., 2002; Makeev et al., 2002; Runborg et al., 2002; Kevrekidis et al., 2003; Siettos et al., 2003; Kevrekidis et al., 2004; Mooller et al., 2005; Moon et al., 2005; Russo et al., 2007]. The underlying assumption of this procedure (see also Figure 1) is that macroscopic models exist and can, in principle, be derived from microscopic ones (description at a much finer level) but they may lack a closed form description (e.g in terms of moments of the microscopic distributions) of their governing-coarse equations. This system identification based computational approach sidesteps the necessity of deriving good *explicit* closures, thus enhancing our efficiency in dealing with the problem in a systematic way.

In [Setayeshgar et al., 2005] it is demonstrated how coarse-projective integration can be carried out using a microscopic-stochastic model of chemotaxis. In [Erban et al., 2006] coarse-projective integration is studied and analysed in detail for kinetic Monte-Carlo of random walks simulating bacterial chemotaxis.

Here we demonstrate how "coarse timesteppers" can be applied to both perform (i) system level steady state and stability analysis and (ii) enable the acceleration of the evolution computations of a Markov chain-Monte Carlo simulation of the Esherichia coli (E.coli) chemotaxis. Through the proposed approach, we show, for the first time, that near equilibrium the coarse-grained dynamics of this system evolve on a two-dimensional slow manifold that can be parametrized by the mean and the variance of the microscopic distribution.

The paper is organized as follows: in section 2 we present the Monte-Carlo model which describes the dynamics of the chemotaxis of E-Coli at micro level. In section 3 we present the proposed framework for multiscale computations. In section 4 we show how the concept of coarse timesteppers can be combined with system identification techniques in order to accelerate in time the detailed simulations. Section 5 provides the results of the analysis and we conclude in section 6.

**2. Esherichia Coli Chemotaxis: A Biased Random Walk Model**

Bacteria, such as E. coli, have several flagella that help them direct their movements towards the most likely direction. Each flagellum rotates by the action of a rotary motor. A typical E-coli bacteria has up to 6 flagella. These can rotate either in a clockwise (CW) manner, causing the bacterium to tumble, or counter-clockwise (CCW), causing the bacterium to swim in a straight line. The overall movement of a bacterium is the result of altering tumble and swim phases. Tumbling frequency and duration time are most important factors of controlling the steering of the bacteria. When a decreasing concentration of an attractant or increasing concentration of a repellant is detected, the bacteria will increase the tumbling frequency and will thus increasing the probability of changing direction.

A bacterium has three types of receptors (transmembrane proteins) for detecting changes in the concentration of attractants and repellents. The signals from these receptors are transmitted through the signal transduction system to the flagella motors, where Che proteins are activated. Increasing concentrations Che-Y protein, which is the output of this system, increases the tumbling frequency, and therefore bias the CW rotation. In general, the evolution of the changes in the transduction system can be represented by an evolution equation of the form:

$$\frac{d\boldsymbol{u}}{dt} = \boldsymbol{F}(\boldsymbol{u}, S) \qquad (1)$$

where $\boldsymbol{u} \in R^n$ is the state vector and $S \in R$ denotes the stimulus signal to the system. In most of the systems, there is some kind of adaptation to the input signal: when $S$ is time independent some functional of $\boldsymbol{u}$ should be also time-independent.

In this paper a simple two dimensional ODEs model for excitation and adaptation is used [Spiro et al., 1997]:

$$\frac{du_1}{dt} = \frac{f(S(t)) - (u_1 + u_2)}{\tau_e} \qquad (2a)$$

$$\frac{du_2}{dt} = \frac{f(S(t)) - u_2}{\tau_a} \qquad (2b)$$

$f$ is the function encoding the signal transduction steps, and should satisfy the condition $f(0) = 0$. The response of this simple model occurs in two time scales: the scale of excitation, which is characterized by the time constant $\tau_e$ and the scale of adaptation, which is characterized by the time constant $\tau_a$.

If $\tau_e \ll \tau_a$, then whenever $t \gg \tau_e, \tau_a$ the model relaxes to

$$u_2 = f(S) - u_1 \qquad (3)$$

The probability that the cell is running (i.e the number of flagella rotating CCW is greater equal to $N/2$, $N$ being number of flagella) is given by the sum of the binomial probability mass function (voting hypothesis):

$$p_{CCW} = \sum_{j=N/2}^{N} \binom{N}{j} p_{CCW}^j (1 - p_{CCW})^{N-j} \qquad (4)$$

From experimental data we know that the probability bias, $p_{CCW}$ of a single flagella motor in the absence of stimulus, for a 2.95 µM concentration of Che-Y protein, is 0.64 [Spiro et al., 1997].

For $N = 6$, eq. (4) gives that the probability of CCW bias of a single bacteria is $p_{CCW} = 0.88$.

The behavior of a single bacterium is determined by the following evolution rules [Spiro et al., 1997]:

1. Draw a random number $\zeta$ from a uniform random distribution on $[0\,1]$.

2. Compare $\zeta$ with the probability of switching direction of rotation:

$$p = 1 - \Phi_\pm \sim k_\pm \pm \Delta t \qquad (5)$$

where $\Delta t$ is the sampling time, $k_\pm$ are rational functions of the reaction rate constants and the concentration. Che-Y. Cluzel, et al. (Science, 2000) have measured the equilibrium CW ($P_{CW}$) and CCW bias ($P_{CCW}$) as a Hill function of the dissociation constant $K_d$ and the concentration $Y$ of Che-Y as follows:

$$P_{CW} = \frac{Y^n}{K_d^n + Y^n} = \frac{k_+}{k_+ + k_-} \qquad (6a)$$

$$P_{CCW} = \frac{Y^n}{K_d^n + Y^n} = \frac{k_+}{k_+ + k_-} \qquad (6b)$$

where $n = 10.3$ and $K_d = 3.1$.

3a. *IF a flagellum rotates CW AND $\zeta > p$ THEN*
   ***keep rotating CW***
   *ELSE*
   ***Switch to CCW rotation***
   *ENDIF*

3b. *IF a flagellum rotates CCW AND $\zeta > p$ THEN*
   ***keep rotating CCW***
   *ELSE*
   ***Switch to CW rotation***
   *ENDIF*

4. IF the number of flagella that rotate CW is less than 3 *THEN*
       *tumble*
   *ELSE*
       *run*
       *IF previously running, THEN*
           *direction remains unchanged*
       *ELSE*
           *direction = +/- 1, with equal probability*
       *ENDIF*
   *ENDIF*

## 3. Multiscale computations of the E-Coli chemotaxis using coarse timestepping

### 3.1 Computational framework for steady-state and stability analysis

We start with a brief overview of the "coarse timestepper". In order to make clear the underlying concept, let us assume that due to the complexity of the system under study there are no explicit macroscopic equations in a closed form that can approximate the emergent macroscopic dynamics in an efficient manner. Yet, we assume that the physics are known in a more detailed level and thus we can develop a "good" microscopic computational model using simulation techniques such as Molecular dynamics, Monte-Carlo, Brownian dynamics and cellular automata. In general, the microscopic simulator can be represented by the following map:

$$U_{k+1} = S_T(U_k, p), \quad (7)$$

where $U_k \equiv U(t_k)$ denotes the state vector of the microscopic distribution at time $t_k = kT$ and $S_T : R^N \times R^m \to R^N$ is the time-evolution operator, $p \in R^m$ is the vector of the system's parameters.

Hence, the microscopic simulator reports the values of the states of the microscopic distribution after an arbitrarily chosen macroscopic time interval $T$.

Let us further assume that the macroscopic (system-level) dynamics of the system under study can be described by a map of the form

$$u_{k+1} = F_T(u_k, p), \quad (8)$$

where $u \in R^n$ denotes the macroscopic state vector, and $F_T : R^n \times R^p$ is a smooth macroscopic time-evolution operator. We furthermore assume that due to complexity of the problem under study such a macroscopic model is not available, or it is overwhelming difficult to derive in a closed form.

When this is the case, a series of important system level tasks such as the tracing of coarse-grained solution branches in the parameter space, stability analysis, optimization and design of control systems cannot be performed by exploiting the arsenal of well-established techniques in the continuum.

The question that naturally arises is how one can systematically study and analyze the macroscopic (system-level) dynamics, when the macroscopic evolution operator is not explicitly available in a closed form. The answer comes from the concept of timestepping which constitutes the "heart" of the Equation-Free computational approach [Kevrekidis et al., 2003, Comm. Math. Science, 1,715; Makeev et al., 2002, J. Chem. Phys., 116, 10083 ; Gear et al., 2002, Comp. Chem. Engng. 26, 941 ; Siettos et al., 2003, J. Chem. Phys., 118, 10149]. The main idea is to sidestep the derivation of system-level equations and to use appropriately-initialized short runs in time of the microscopic/stochastic models to estimate necessary quantities "on demand". What the Equation-Free approach does, in fact, is providing closures "on demand"; relatively short bursts of the fine scale simulator naturally establish in a strictly numerical manner the slaving relation between the fast and slow dynamics of the system under study [refer to Gear et al., 2002, Comp. Chem. Engng. 26, 941; Siettos et al., 2003, J. Chem. Phys., 118, 10149; Kevrekidis et al., 2003, Comm. Math. Science, 1, 715 for more detailed discussions].

In a nutchel, the coarse timestepper consists of the following steps (see also figure 1):

(a) Prescribe a macroscopic initial condition (e.g. concentration profile) $u(t_0)$;

(b) Transform it through lifting to one (or more) fine, *consistent* microscopic realizations $U(t_0) = \mu\, u(t_0)$, where $\mu$ is the lifting operator.

(c) Evolve this(ese) realization(s) using the microscopic simulator for the desired short macroscopic time (time horizon) $T$, generating the value(s) $U(T)$. An appropriate choice $T$ can be estimated from the spectrum derived of the mean field equations linearization around their steady states.

(d) Obtain the restrictions $u(T) = M\, U(T)$; $M$ denotes the restrict operator. The lift and restrict operators should satisfy $\mu M = I$.

The above procedure can be considered as a "black box" coarse timestepper

$$u_{k+1} = \Phi_T(u_k, p) \qquad (9)$$

Under certain assumptions regarding the separation between the time-scales of the system [Kevrekidis et al. 2003], $\Phi_T : R^n x R^p \to R^n$ provides an numerical approximation of the unavailable $F_T$.

Thus:

(e) At an outer level, and depending on the task we want to carry out, (such as the computation of fixed points, the stability analysis and optimisation), well established numerical analysis algorithms can be utilized to *estimate* "on demand" the required quantities such as residuals, Jacobians, control matrices and Hessians. These algorithms call the timestepper as a black-box subroutine from nearby appropriately perturbed initial conditions and for relatively short time intervals.

For example, coarse-grained (macroscopic) steady states can be obtained as fixed points, using $T$ as sampling time, of the mapping

$$\Phi_T : u - \Phi_T(u, p) = 0 \qquad (10)$$

using Newton-Raphson method [Kelley, 1995]. The procedure involves the iterative solution of the following linearized system:

$$\left[ I - \frac{\partial \Phi_T}{\partial u} \right] \delta u = -[u - \Phi_T(u, p)] \qquad (11)$$

The computation of the Jacobian $\frac{\partial \boldsymbol{\Phi}_T}{\partial \boldsymbol{u}}$ can be achieved in a fully numerically manner by calling the coarse timestepper from appropriately perturbed initial conditions. For example, using center finite differences the $(i, j)$ element of the Jacobian is given by:

$$\frac{\partial \Phi_{T\,j}}{\partial u_i} = \frac{\Phi_{T\,j}(u_i + \varepsilon(u_i)) - \Phi_{T\,j}(u_i - \varepsilon(u_i))}{2\varepsilon(u_i)} + O(\varepsilon(u_i)^2) \quad (12)$$

where $\Phi_{T\,j}$, $u_i$ denote the $j-th$ and $i-th$ elements of $\boldsymbol{\Phi}_T$ and $\boldsymbol{u}$ respectively; $\varepsilon$ is a small and appropriately chosen scalar.

The above framework enables the temporal simulator to converge to both stable and unstable solutions and trace their locations through bifurcation points utilizing techniques such as the pseudo-arc length continuation approach.

For large-scale systems, fixed point calculations can be performed in a more efficient manner using matrix-free iterative solvers such as the Newton-Generalized Minimum Residual (Newton-GMRES) method [Kelley, 1995]. The advantage of using matrix-free methods over more "traditional" techniques is that the explicit calculation and storage of the Jacobian $\frac{\partial \boldsymbol{\Phi}_T}{\partial \boldsymbol{u}}$ is not required. Instead, what is really needed is matrix-vector multiplications which can be obtained at low computational cost by calling the timestepper from *nearby* initial conditions allowing the estimation of the action of the linearization of the map $\boldsymbol{\Phi}_T$ on known vectors, as

$$D\boldsymbol{\Phi}_T(\boldsymbol{u}) \cdot \boldsymbol{v} \approx \frac{\boldsymbol{\Phi}_T(\boldsymbol{u}+\varepsilon \boldsymbol{v}) - \boldsymbol{\Phi}_T(\boldsymbol{u})}{\varepsilon} \quad (13)$$

Alternative algorithms such as the Recursive Projection method [Shroff & Keller, 1993] and other Newton-Picard methods such as the ones presented in [Lust et al., 1998; Kavousanakis et al., 2008] can be also used to compute both steady states and periodic solutions and construct their bifurcation diagrams.

The leading (algebraically largest) eigenvalues of the matrix $D\boldsymbol{\Phi}_T$ determine the local stability of the system. For large-scale systems these can be estimated using again a matrix-free iterative eigensolver such as the Arnoldi procedure [Saad, 1992; Christodoulou & Scriven 1998] exploiting the same timestepper approach. These algorithms share common procedures that in short can be described as follows (see also figure 2):

Choose $\boldsymbol{v}_1 \in R^n$ with $\|\boldsymbol{v}_1\| = 1$
*For j =1 Until Convergence*
   (1) Compute and store $D\boldsymbol{F}_T \cdot \boldsymbol{v}_j$ using (9)
   (2) Compute and store $h_{t,j} = \langle D\boldsymbol{\Phi}_T \cdot \boldsymbol{v}_j, \boldsymbol{v}_t \rangle, t = 1,2,\ldots,j$
   (3) $r_j = D\boldsymbol{\Phi}_T \cdot \boldsymbol{v}_j - \sum_{t=1}^{j} h_{t,j} \boldsymbol{v}_t$
   (4) $h_{j+1,j} = \langle r_j, r_j \rangle^{1/2}$
   (5) $\boldsymbol{v}_{j+1} = r_j / h_{j+1,j}$
End

At step $j$, the algorithm produces an orthonormal basis $\{v_1, v_2,...., v_m\}$ of the Krylov subspace $K_j$ spanned by

$$\{v_1, D\Phi_T \cdot v_1,...., D\Phi_T^{j-1} \cdot v_{j-1}\} \quad (14)$$

The projection of $DF_T$ on $K_j$ is represented in the basis $\{v_j\}$ by the upper Hessenberg matrix

$$H_j = v_j^T D\Phi_T \cdot v_j \quad (15)$$

whose elements are the coefficients $h_{ij}$.

At each $j$ step, GMRES minimizes the residual $R = \|u - \Phi_T(u, p)\|$.

Regarding the stability analysis, the eigenvalues of $H_j$ provide approximations of the $D\Phi_T$ for the outermost spectrum of $D\Phi_T$, where the eigenvectors of $D\Phi_T$ are approximated by

$$x_j = v_j z_j, \quad (16)$$

Where $z_j$ are the eigenvectors of the Hessenberg (it may be computed using standard packages like EISPACK or LAPACK).

The approximation character of the algorithm, flashes a note of caution in the choice of the perturbation parameter $\varepsilon$ since at the end of the algorithm what is usual done is the formation of the converged Hessenberg through the multiplication

$$H_j = v_j^T (Av_j) \approx v_j^T \frac{\Phi_T(u + \varepsilon v_j) - \Phi_T(u)}{\varepsilon} \quad (17)$$

which is only an approximation to the actual Hessenberg. This may reflect to some "distortion" to the computation of the leading eigenvectors. One should choose $\varepsilon$ according to the accuracy of the already converged solution.

**3.2 The lift and restrict operators for the chemotaxis problem**

As explained in 3.1, we don't try to find any closures. Instead we exploit the concept of coarse timestepping to sidestep the derivation of such approximations. For our analysis, the spatial distribution in $x$, $\rho(x)$ is computed by approximating the corresponding inverse cumulative distribution function $ICDF(\rho)$ using for example $n_s$ orthogonal polynomials or splines of order $q$ (see also figure 2). For example let us denote by $W$ the matrix of dimension $(m \times n_s)$ containing the values of such $n_s$ orthogonal polynomials over the $m$ points in space on which the $ICDF(\rho)$ is computed [Setayeshgar et al., 2005].

The restrict operator is defined as the product of $ICDF(\rho)$ with $W$ resulting to the coefficients of the orthogonal polynomials $a$ i,.e.:

$$a = ICDF(\rho) \cdot W \tag{18}$$

since $WW^T = I$ the lift operator creates $ICDF(\rho)$ (i.e. a distribution in space) as:

$$ICDF(\rho) = a \cdot W^T \tag{19}$$

**4. Coarse Projective Integration of the Markov Chain Monte Carlo Chemotaxis Model.**

The coarse timestepper can be also used to perform coarse projective integration (see Fig. 3). The basic idea is that coarse time-stepper can be used to approximate the time derivatives of the corresponding continuum formulation, even if the continuum equations are not known in closed form. Specifically, we execute the following steps:

> **(f)** Repeat step **(d)** over several time steps, giving several $U(t_i)$, as well as their restrictions $u(t_i) = M\,U(t_i), i = 1, 2, ...., k+1$.
> **(g)** Use the chord connecting these successive time-stepper output points to estimate the derivative of the continuum variables. Note that this does *not* require that we know the explicit continuum equations.
> **(h)** Use this derivative in an outer integrator (such as forward Euler) to estimate the continuum state $u(t_{k+1+m})$ much later in time.
> **(i)** Go back to step **(b)**.

For the chemotaxis problem, the proposed procedure can be summarized in a two-tier level pseudo-code as follows:

> ***Do while {desired time $T_{final}$}***
>
> **I. For $t = n, n+1, n+2, \ldots n+l$**
> (a)  Compute the Cumulative distribution function of x-positions $CDF(\rho)_t$ from the probability distribution function $PDF(\rho)_t$
> (b)  Approximate the Inverse Cumulative Distribution function of *x*-positions $ICDF(\rho)_t$ using $n_s$ orthogonal polynomials. Figure 4 shows five such orthogonal polynomials.
> (c)  Restore the coefficients $a = [a_1, a_2, ..., \alpha_{ns}]_t$ of the approximating polynomials.
> ***End For***
>
> **II. Project the polynomial coefficients k-times instances ahead to estimate the $ICDF(\rho)_{t+l+k}$.**
> Several techniques can be used for that purpose, including ARMAX models [Ljung, 1987] and polynomial curve fitting using Least Squares; At this point we can also incorporate discrete time filters to smooth the time-series signal. This step involves the selection of an appropriate model structure (among a set of candidate models). Here, we demonstrate an Autoregressive (AR)-based system identification technique:

(a) Perform system identification of the approximating polynomial coefficients of the ICDF by a discrete model of the form

$$a_i(t) = c_1 a_i(t-1) + c_2 a_i(t-2) + \ldots + c_{n-k} a_i(t-n_k), \quad i = 1, 2, \ldots, n_s$$

The above expression relates $a_i$ at time t to a finite number of past values $a_i(t-n_k)$. At this point we should note that, if necessary, a smoothing filter can be applied to the $a_i$ time series.

In this work, we implemented a second order (in a window of 5 sample points), Savitzky Golay filter [Savitzky and Golay, 1964], which is a time varying FIR filter. Savitzky-Golay smoothing filters (also called digital smoothing polynomial filters or least squares smoothing filters) are typically used to "smooth out" a noisy signal whose frequency span (without noise) is large. Savitzky-Golay filters are optimal in the sense that they minimize the least-squares error in fitting a polynomial to each frame of noisy data. The estimation of the vector of parameters $\boldsymbol{\Theta}^T = [c_1, \ldots, c_{n-k}]$ at time t is carried out using least squares (LS) over the "raw" or smoothed data at time instances $t = n, n+1, \ldots n+l$.

(b) Use the derived model to project $a_i$ to time $t+l+k$.
(c) Use the projected values to lift to one (or more) consistent microscopic distributions of positions $x$

*End While*

It is important to note that one must integrate the microscopic rules for some time before estimating the time derivative of the continuum variables. This allows higher moments of the continuum description to become slaved to the statistics of interest.

## 5. Numerical Analysis

Simulation results were obtained using 2000 cells ($N_{cells}$) and $dt = 0.1$ as the Monte Carlo time step. The chemoatractant profile is a Gaussian-like function given by $S = \frac{1}{\sqrt{\pi}} e^{-\frac{(x-6)^2}{2}}$ (Figure 5a).

Figure 5b illustrates the left and right moving distributions for the Gaussian attractant profile. The distributions were derived by averaging over time from $t=10000$ *s* to $t=15000$ *time steps* using a time horizon of $T=100$ *time steps* (i.e averaging over a total of 50 samples) and over the total distribution of positions. The initial condition was a uniform (flat) profile of positions. As it can clearly shown, simulation results are consistent with those dictated by the voting hypothesis giving a probability of ~ 0.88 that the cell is running either left or right.

Figure 6 depicts the distributions of the number of left moving flagella that rotate CCW (figure 6a) and tumbling cells (Figure 6b). Simulation results are consistent with those dictated by the binomial probability mass functions.

For the problem under study the analysis is based on the hypothesis that a macroscopic coarse model exists and closes for the microscopic distribution of bacteria positions $\rho(x)$. This implies that all the other "inner" variables, such as number of flagellea rotating CCW or CW, the excitation $u_1$ and adaptation $u_2$ signals, become quickly slaved to the spatial distribution (they evolve towards a "slow manifold" parameterized by the underlying microscopic distribution). Computational results corroborating this can be seen in Fig 7, which illustrates the relatively fast slaving of the "internal" variables. Here we initialized the bacteria by setting $u_1 = u_2 = 0$ for all cells, running right with 4 flagellae rotating CCW.

For the coarse-projective integration, the approximation of *ICDF* is made with linear extrapolation in time, $l = 10$ and $k = 10$ using eight orthogonal basis functions until $t = 6000$ *time steps* and then $l = 10$, $k = 20$ till $t = 25000$ *time steps*, while after each lifting a period of $5T$ is used as "healing" of the errors made in the lifting step before the acquisition of the training data. In figure 8, are given the evolution of both normal integration (blue lines) and coarse-projective integration (red lines) for the density and cumulative distribution functions.

The steady-state and stability analysis was performed using two different methods, namely (i) through Newton-Raphson's method and (ii) through matrix-free methods.

Newton-Raphson's method was applied in terms of the mean value and the variance of a normal distribution function. Under this assumption, lifting was done with the inverse normal distribution. The fixed point was evaluated through the coarse timestepper with $T = 400$ *time steps*, 10 copies of cell distributions. Upon convergence of the Newton-Raphson to a residual of $O(10^{-3})$, for perturbations $\varepsilon \sim 5\times 10^{-2}$) the mean value was found to be $\sim 6.005$ and the final variance $\sim 0.13$. The values of the mean and standard deviation of the normal distribution in each step of Newton's iteration is shown in figure 9b. The initial values of the mean and standard deviation of the distribution at $t = 10000$ *time steps* where we started Newton are also shown. Newton-Raphson's method converged in 3 iterations (see figure 10a). Upon convergence, the Jacobian was found to be $A = \begin{bmatrix} 0.90521 & -0.0010 \\ -0.012 & 0.8423 \end{bmatrix}$. The eigenvalues of the 2x2 matrix are: $\lambda_1 \sim 0.91$ and $\lambda_2 \sim 0.845$. Hence the corresponding eigenvalues in the s-plane are: $\lambda_1 = -0.023$ and $\lambda_2 = -0.0421$. Looking at the Jacobian, the interaction of the non-diagonal terms is very small when compared with the diagonal terms. This is an indication that near the fixed point the particular system can be considered for practical purposes uncoupled.

Newton-GMRES was used to compute the steady states with respect to the coefficients of the orthogonal basis functions and Arnoldi's method has been employed to perform stability analysis. These matrix-free algorithms were wrapped around the microscopic simulator with $T = 2000$ *time steps*, averaging over 20 copies of cells distributions. For the approximation of the inverse cumulative distribution function we used eight orthogonal basis functions and we asked for the 2-leading eigenpairs with a residual of $O(10^{-4})$.

For comparison purposes the eigenvalues (in the s-plane) as calculated using Newton-Raphson's method with lifting through the inverse normal distribution function and the Arnoldi's method using eight orthogonal basis functions are given in table 1.

Figure 10b shows the eigenvectors corresponding to the symmetric (variance of distribution of bacteria positions) and antisymmetric (mean of the distribution of bacteria positions) part of perturbation.

As it is shown, the leading eigenvalues and their corresponding eigenvectors as computed with Newton-Raphson and Arnoldi coincide for any practical means. This reveals that near the equilibrium the coarse-grained dynamics evolve on a two dimensional manifold that can be in principle parameterized by the first two moments of the underlying microscopic distribution, namely the mean and the variance of the spatial distribution.

## 6. Conclusions

The Equation-free approach is a computational framework that provides a systematic approach for analyzing the parametric behavior of complex/ multiscale simulators much more efficiently than simply simulating forward in time. Acceleration of simulations in time, regime and stability computations, as well as continuation and numerical bifurcation analysis and other important tasks such as rare-events analysis of the complex-emergent dynamics can be performed in a straightforward manner bypassing the explicit extraction of closures.

In this work, we have demonstrated how this multiscale approach can be used to extract system-level information from a Monte-Carlo model simulating the chemotaxis of E. COLI. Fixed point and stability analysis were performed using both "conventional" contraction maps such as Newton-Raphson and matrix-free iterative algorithms (Newton-GMRES for fixed point computations and Arnoldi eigensolver). Through the numerical analysis we showed that near the equilibrium the coarse-grained dynamics evolve on a two dimensional manifold that can be defined as a function of the first two moments of the evolving spatial distribution. Coarse projective integration was also demonstrated combining the concept of coarse-timestepping with nonlinear system identification techniques. The particular one-dimensional in space stochastic model does not exhibit any striking nonlinear behavior such as the appearance of critical points, phase transitions or blow-up solutions in finite time. Such phenomena including complex pattern formation have been observed in experimental-observed chemotactic responses [Adler and Templeton, 1967; Budrene and Berg, 1991, 1995] and their dynamics have been approximated and analysed using macroscopic level models [Keller and Segel, 1971a,b; Brenner et al., 1998; Myerscough et al., 1998; Polezhaev et al., 2006] . To this end we believe that the proposed framework can be used to deepen our understanding in the way chemotaxis causes the emergent of such complex behaviour. Systematic coarse-grained bifurcation and stability analysis for both steady state and periodically oscillatory solutions [Russo et al., 2007; Kavousanakis et. al., 2008], the adaptive detection of the critical points that mark the onset of such phenomena [Siettos et al., 2006] can be efficiently performed exploiting the detailed biological knowledge that is incorporated into state-of-the art-microscopic models.

**Acknowledgments**

The author would like to thank Professor Yannis Kevrekidis (Dept. of Chemical Engineering, Dept. of Mathematics and PACM, Princeton University), Professor C. W. Gear (Dept. of Chemical Engineering, Princeton University) and Professor Hans Othmer (Dept. of Mathematics, University of Minnesota) for their collaboration, insightful comments and many fruitful and helpful discussions.

**Figures**

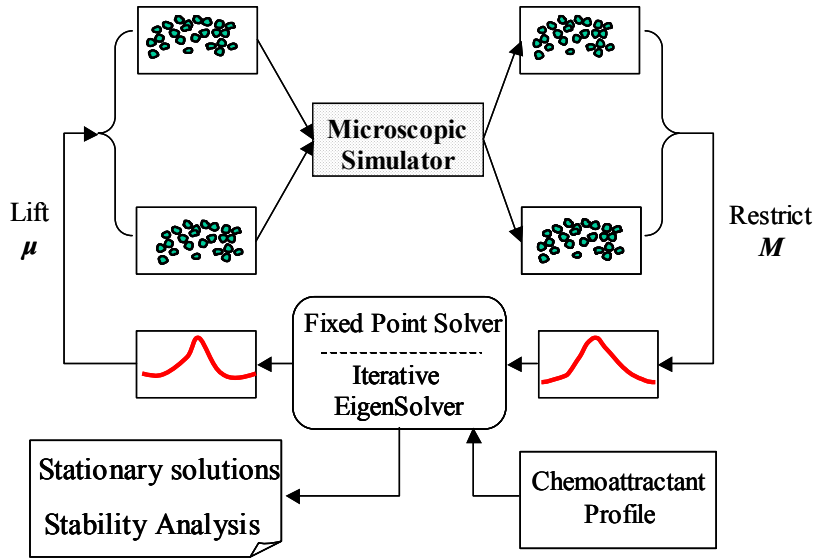

**Figure 1:** The coarse timestepper

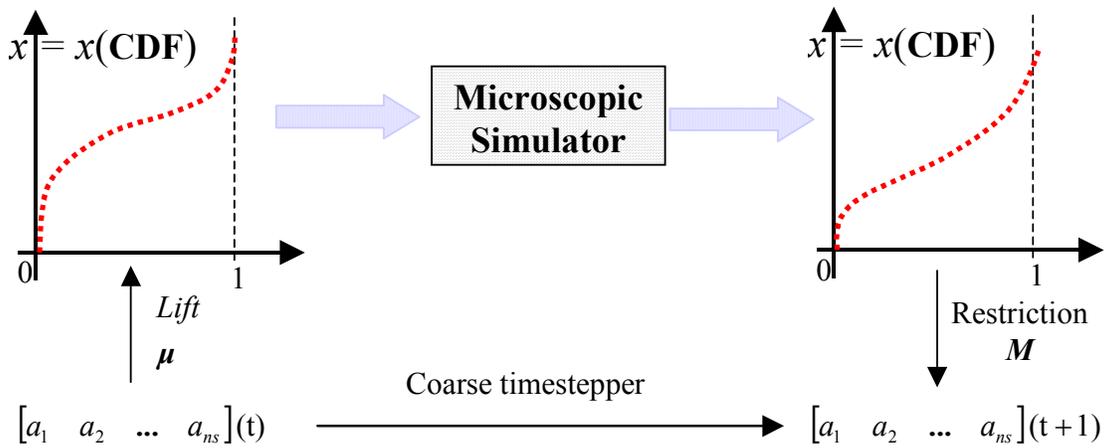

**Figure 2.** Lifting and Restriction of the microscopic distribution

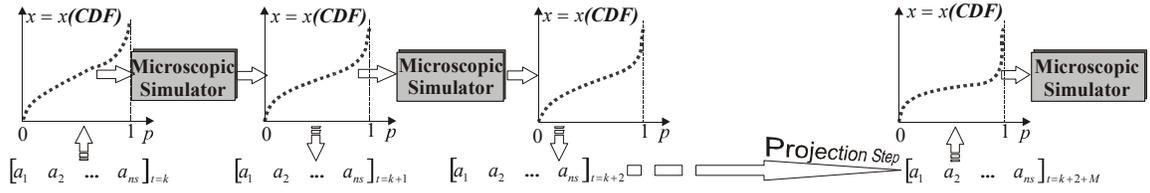

**Figure 3:** Coarse projection

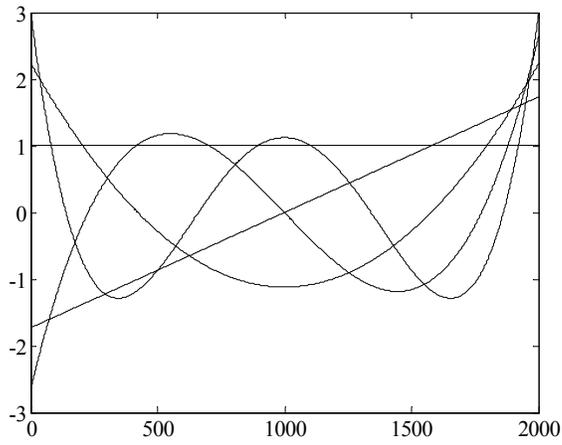

**Figure 4**. Orthogonal basis functions used to approximate the ICDF.

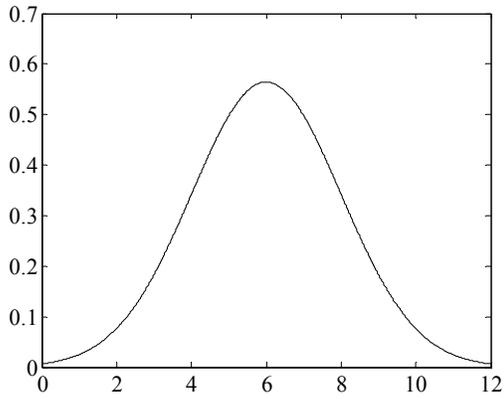 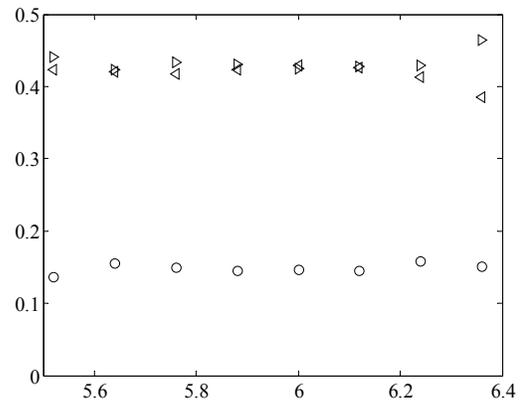

(a) (b)

**Figure 5**. (a) Gaussian attractant profile, (b) Distributions of bacteria: Running left < , right > and tumble o.

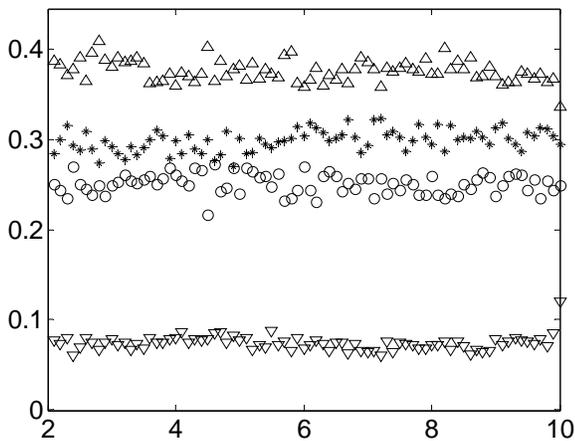 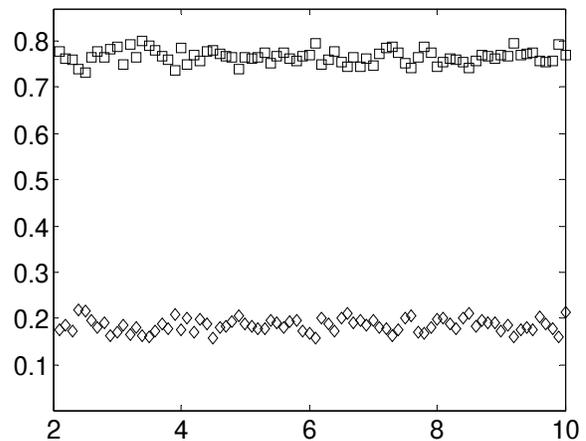

(a)

**Figure 6**. (a) Flagella rotating CCW Distributions for the left moving cells; triangles-up correspond to 4, asterics to 3, circles to 5 and triangles-down to 6 CCW rotating flagella (b) Flagella rotating CCW Distributions for the tumbling cells-Flat attractant distribution. Squares correspond to 2 and rhombs to 1 CW rotating flagella.

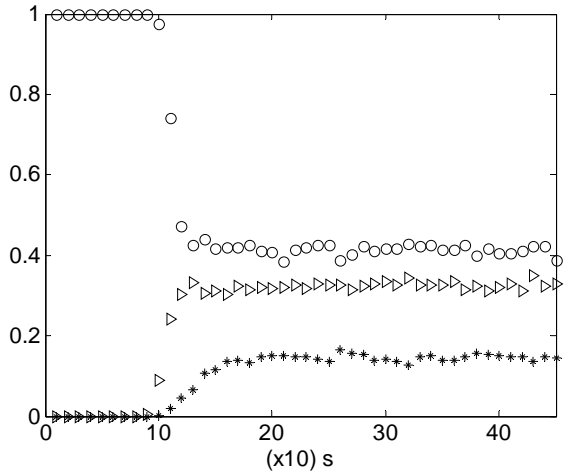 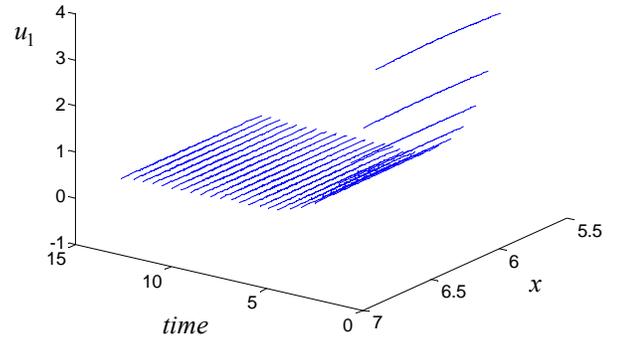

(a)                          (b)

**Figure 7.** Quick dynamic slaving of "internal" variables. When lifting, (i) we set u(1) = u(2)= 0 for everybody and (ii) we set all cells running with 4 flagellae CCW (iii) we start with all cells running RIGHT. (a) Time evolution of number of cells running right (circles), running cells (either left or right) having 4 flagella rotating CCW (triangles right), tumbling cells (asterisks), (b) evolution of variable $u_1$. (Gaussian attractant profile)

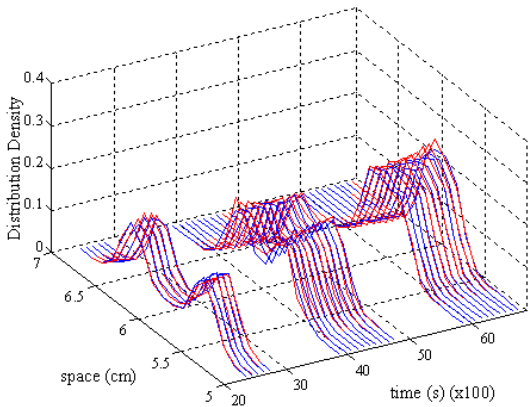 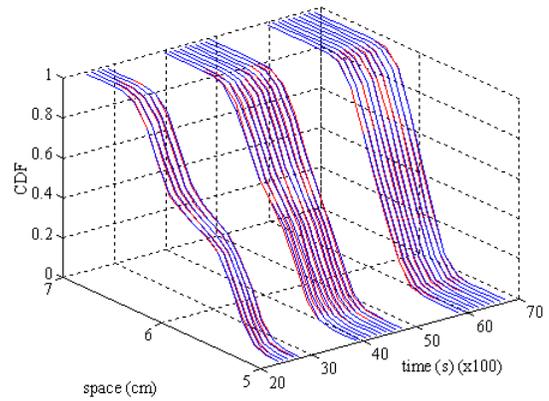

(a)                          (b)

**Figure 8**. Coarse projection (5 healing, m=5 acquisition, k=10 projection & 8 basis functions till time=6000 and then 5 healing, 5 acquisition, k=30 projection & 8 basis functions till time=25000). Blue lines correspond to temporal simulations while red ones to coarse projected simulation.

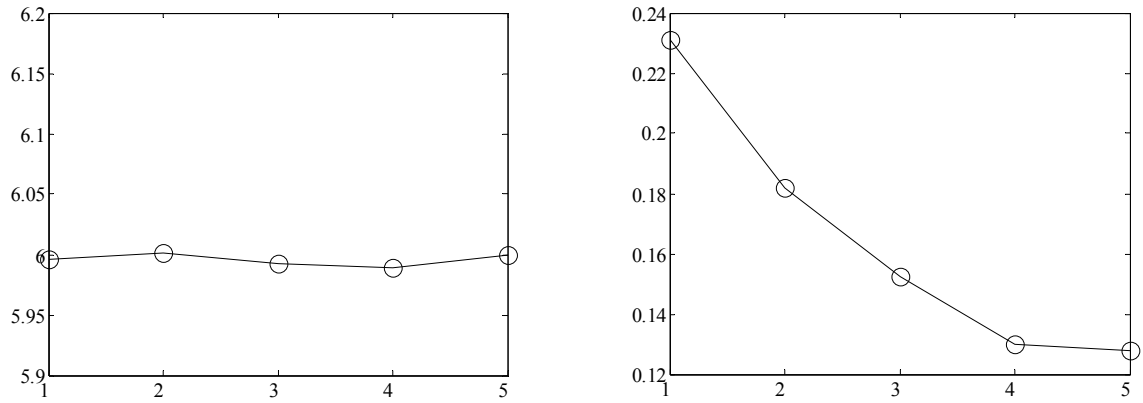

**Figure 9**. (a) Mean and (b) standard deviation of the normal distribution in each step of Newton's iteration.

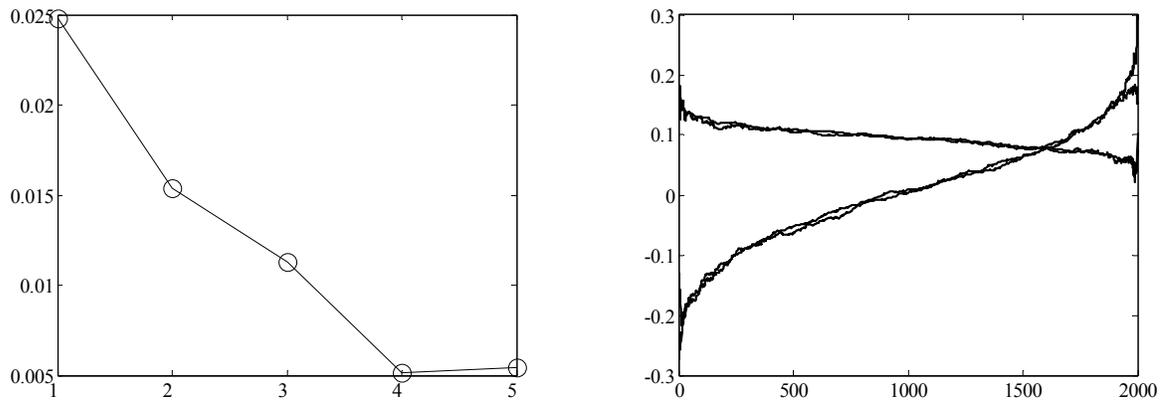

**Figure 10**. (a) Newton's convergence (error-iterations) (b) System Eigenvectors of the ICDF; solid lines correspond to eigenvectors as computed with Newton's method while the dotted ones correspond to the eigenvalues as computed with Arnoldi's methos; from left: upper (bottom) lines correspond to $\lambda_1 \sim 0.61$ ($\lambda_2 \sim 0.44$).

| ARNOLDI | NEWTON-RAPHSON |
|---|---|
| $\lambda_1 = -0.025$ | $\lambda_1 = -0.023$ |
| $\lambda_2 = -0.041$ | $\lambda_2 = -0.042$ |

Table. 1. **Eigenvalues as calculated using Newton's method with lifting through the inverse normal distribution function and the Arnoldi's method using 5 orthogonal basis functions.**